\documentclass[onecolumn,aps,12pt]{revtex4}
\usepackage{amsmath,graphicx}
\usepackage{amsmath,graphicx} 
\def\,{\ifmmode\mskip\thinmuskip\else\leavevmode\thinspace\fi}

\newcommand{\dd}{\mbox{d}}
\newcommand\ba{\begin{eqnarray}}
\newcommand\ea{\end{eqnarray}}

\def\fun#1#2{\lower3.6pt\vbox{\baselineskip0pt\lineskip.9pt
\ialign{$\mathsurround=0pt#1\hfil##\hfil$\crcr#2\crcr\sim\crcr}}}

\begin{document}

\title{ Compton and double Compton scattering 
processes at colliding electron-photon beams }
\author{A.N. Ilyichev}
\affiliation{NC PHEP 220040 Minsk, Belarus}
\author{E.A. Kuraev, V. Bytev}
\affiliation{Joint Institute for Nuclear Research, Dubna, 141980, Russia}
\author{Yu. P. Peresun'ko}
\affiliation{NSC KIPT, Institute of Theoretical Physics, Kharkov,
Ukraine}


\begin{abstract}
Radiative corrections (RC) to the Compton scattering cross section are calculated
in the leading and next-to leading logarithmic approximation to the case of
colliding high energy photon-electron beams.
 
RC to the double Compton scattering cross section in the same experimental set-up
are calculated in the leading logarithmic approximation. 

We consider the case when no pairs are created in the final state.
We show that the differential cross section can be written
in the form of the Drell-Yan process cross-section.
Numerical values of the $K$-factor and the leading order distribution on the
scattered electron energy fraction and scattering angle are presented.
\end{abstract}
\maketitle

\section{Introduction}
The Compton scattering process 
\begin{gather}
\gamma(k_1)+e^-(p_1) \to \gamma(k_2)+e^-(p_2),\nonumber \\
k_1^2=k_2=0,\quad p^2_1=p^{2}_2=m^2, \\  \nonumber
\kappa_1=2p_1k_1=4\epsilon_1\omega_1,\quad \kappa_1^{'}=2p_2k_1=
2\epsilon_2\omega_1(1-c), \quad s_1=2p_1p_2=2\epsilon_1\epsilon_2(1+c), \\ \nonumber
\kappa_1\sim\kappa_1^{'}\sim s_1\gg m^2, \quad
\epsilon_2=\frac{2\epsilon_1\omega_1}{\omega_1(1-c)+\epsilon_1(1+c)},
\end{gather}
(with $\epsilon_{1,2},\omega_1$-are the energies of initial and 
 scattered electrons and the initial photon; $c=\cos\theta $, $\theta $ is
the angle between $\vec{p}_2,\vec{k}_1$)-
plays an important role as a perspective calibration process 
at high-energy photon-electron colliders \cite{exp}. To 
obtain a radiative corrected cross section of this process is the 
motivation of this paper. Modern methods based on the renormalization 
group approach in combination with the lowest order radiative corrections
(RC) permit us to obtain a differential cross section with
the leading ($((\alpha/\pi)L)^n\sim 1$, with "large logarithm" 
$L=\ln(s_1/m^2)$) and the 
next-to leading approximation (i.e. keeping the terms of the order 
$(\alpha/\pi)^nL^{n-1}$). So the  accuracy of the formulae given 
below
is determined by the terms of the order (which are systematically omitted)
\ba
\frac{m^2}{\kappa_1},\quad \frac{\alpha^2}{\pi^2}L,\quad
\alpha\frac{s}{M_Z^2},
\ea
compared with the terms of order  of unity and is at the level of per-mille
for typical experimental conditions \cite{exp} 
$ \theta\sim 1,\kappa_1 \sim 10 GeV^2$. We consider the energies of 
initial particles to be  much less than the Z-boson mass $M_Z$ and, therefore,
the weak corrections to the Compton effect in our consideration 
are beyond  our accuracy.

The first papers devoted to cancellation of radiative corrections
to Compton scattering were published  in 1952 by  Brown and
Feynman \cite{BF} (virtual and soft real photon emission contribution),
and Mandl and Skyrme \cite{MS} (emission of an additional hard photon).

In the work of H. Veltman \cite{polcompt1}, the lowest order radiative corrections
to the polarized Compton scattering were calculated in non-relativistic
kinematics. This case of kinematics was
also considered in the paper of M. Swartz \cite{polcompt2} .

In the papers of A. Denner and S. Dittmaier \cite{polcompt3},
the lowest order radiative corrections  in the framework of Standard Model
was calculated for the case of polarized  electron and photon.

In this paper, we consider the case of high-energy electron and photon Compton 
scattering (cms energy supposed to be  much higher
than the electron mass but much less than the Z-bozon mass). 
We found that the cross-section with radiative corrections of all orders
of PT taken into account  could be written down in the form  of the Drell-Yan process.
Both leading  and next-to-leading  contributions are derived explicitly.

We imply the kinematics when the initial photon and electron move along the 
$z$ axis in the opposite directions 
The energy of the scattered electron will be the function of its scattering 
angle:
\ba
z_0=\frac{\varepsilon_2}{\omega_1}
=\frac{2\rho}{a},\quad a=a(c,\rho)=1-c+\rho(1+c), \quad
\rho=\frac{\varepsilon_1}{\omega_1}.
\ea
Hereafter  we imply the kinematic case $\rho<1$.
The case $\rho>1$ will be considered in section \ref{secK}.

The differential cross section in the Born approximation will be
\ba
\label{born}
\frac{\dd\sigma_B}{\dd c}(p_1,\theta)
=\frac{\pi\alpha^2U_{0}}{\omega_1^2 a^2},\quad U_0=\frac{a}{1-c}+\frac{1-c}{a}.
\ea
When taking into account RC of higher orders (arising from both emission 
of virtual and real photons) a simple relation between the scattered 
electron energy and the scattering angle changes, so the differential cross 
section is in general dependent on the energy fraction $z$ of the
scattered electron. Accepting the Drell-Yan form a of cross section, we can put
it in the form
\ba
\frac{\dd\sigma}{\dd c \dd z}(p_1,\theta,z)
=\int\limits_0^1 \dd x D(x,L)\int\limits_z^{z_0}\frac{\dd 
t}{t}
D(\frac{z}{t},L)\frac{\dd\sigma_h}{\dd c \dd t}
(xp_1,\theta,t)(1+\frac{\alpha}{\pi}K),
\ea
where the structure function $D(x,L)$ (specified below) describes the 
probability to find
the electron (considered as a parton) inside  the electron, $K$ is the so-called 
K-factor which can be calculated from the lowest RC orders, and the 
"hard" cross section is
\begin{gather}
\label{bornx}
\frac{\dd\sigma_h}{\dd c \dd t}(xp_1,\theta,t)
=\frac{\dd\sigma_B}{\dd c}(xp_1,\theta)\delta(t-t(x)),
\\ \nonumber
\frac{\dd\sigma_B(xp_1,\theta)}{\dd c}=\frac{\pi\alpha^2}{\omega_1^2}\frac{1}{(1-c+\rho x(1+c))^2}
\biggl(\frac{1-c}{1-c+\rho x(1+c)}+\frac{1-c+\rho x(1+c)}{1-c}\biggr),
\\ \nonumber
t(x)=\frac{2x\rho}{1-c+\rho x(1+c)},
\end{gather}
with $K$ specified below (see (\ref{master},\ref{contrKsv},\ref{contrKh})).

The cross section written in the Drell-Yan form explicitly satisfies the 
Kinoshita-Lee-Nauenberg theorem \cite{KLN}. Really, being integrated 
on the scattered
electron energy fraction $z$, the structure function corresponding to the
scattered electron turns to unity due to its property
\ba
\int\limits_0^1\dd z\int\limits_z^1\frac{\dd t}{t}D(\frac{z}{t},L)f(t)=
\int_0^1\dd t f(t).
\ea
Mass singularities associated with the initial lepton structure function remain.

So our master formula for the cross section with RC taken into account
is 
\begin{gather}
\label{master}
\frac{\dd\sigma}{\dd z \dd c}(p_1,p_2)
=\int\limits_{x_0}^1\frac{\dd x}{t(x)}D(x,L)
\frac{\dd\sigma_B}{\dd c}(xp_1,\theta)D(\frac{z}{t(x)},L)+
\frac{\alpha}{\pi}\frac{\dd\sigma_B(p_1,\theta)}{\dd c}\biggl[K_{SV}\delta(z-z_0)+K_h\biggr],\\ \nonumber
z=\frac{\epsilon_2^{'}}{\omega_1}<z_0,\quad x_0=\frac{z(1-c)}{\rho(2-z(1+c))}, 
\quad L=\ln\frac{2\omega_1^2 z_0\rho(1+c)}{m^2},
\end{gather}
with the nonsinglet structure function $D$ defined as 
\cite{JSW}
\begin{gather}\label{8,
iterative form EST} D(z,L) = \delta(1-z) +
\sum_{k=1}^{\infty}\frac{1}{k!}
\biggl(\frac{\alpha\,L}{2\pi}\biggr)^kP_1(z)^{\otimes k}\ , \\ \nonumber
P_1(z)^{\otimes k}=\underbrace{ P_1(z)\otimes\cdots\otimes
P_1(z)}_{k} \ , \quad
P_1(z)\otimes P_1(z) = \int\limits_{
z}^{1}P_1(t)P_1\biggl(\frac{z}{t} \biggr)\frac{dt}{t} \ ,
\\ \nonumber
P_1(z) = \frac{1+z^2}{1-z}\theta(1-z-\Delta) +
\delta(1-z)\bigl(2\ln{\Delta}+\frac{3}{2}\bigr) \ , \ \Delta \ll 1 \ .
\end{gather}
In Conclusion (see \ref{smoothed}) we put the so-called "smoothed" form of
a structure function.

The second term in rhs of (8) collects all the nonleading contributions 
from virtual, soft, and hard 
photons emission, with $K_{SV}$ given in section \ref{Ksv} where
the virtual and soft real contributions are considered.
In sections \ref{Kcoll}, \ref{conlus},
we consider the contribution from an additional hard photon emission and
introduce an auxiliary parameter $\theta_0$ to distinguish 
the collinear and noncollinear kinematics of photon emission.
Also, we put the expression for the hard photon contribution $K_h$.
The results of numerical estimation for K-factor and leading contributions are
given in section \ref{conc}. In Section \ref{appA} (App. A), we demonstrate the explicit 
cancellation of $\theta_0$ dependence. In section \ref{secK} (App. B), we consider 
the kinematic case $\varepsilon_1>\omega_1$.

\section{Contribution of virtual and soft real photons}
\label{Ksv}

To obtain the explicit form of the $K$ factor, we reproduce the lowest order RC
calculation. It consists of virtual photon emission
contribution and the contribution from the real (soft and hard) 
photon emission taken into account. The virtual and soft  photon  emission 
contribution was first calculated in the famous paper of the 1952 year by Laura Brown 
and Richard Feynman \cite{BF}. The result has the form
\ba
\frac{\dd\sigma_{virt}}{\dd\sigma_B}=-\frac{\alpha}{\pi}\frac{U_1}{U_0},
\ea
with (see \cite{BF}, kinematic case II):
\ba
\frac{U_1}{U_0}=(1-L)(\frac{3}{2}+2\ln\frac{\lambda}{m})+\frac{1}{2}L^2
-\frac{\pi^2}{6}-K_V, \quad
U_0=\frac{\kappa_2}{\kappa_1}+\frac{\kappa_1}{\kappa_2},
\ea
with $K_V$ (virtual photon contribution to the $K$-factor):
\ba
K_V=-\frac{1}{U_0}\biggl[(1-\frac{\kappa_2}{2\kappa_1}-\frac{\kappa_1}{\kappa_2})
(\ln^2\frac{s_1}{\kappa_1}-\ln\frac{s_1}{\kappa_1}
+2\ln\frac {\kappa_2}{\kappa_1})\\
+(1-\frac{\kappa_1}{2\kappa _2}-\frac{\kappa _2}{\kappa_1})
(\ln^2\frac{s_1}{\kappa_2}-\ln \frac{s_1}{\kappa_1}
-\ln\frac {\kappa_1}{\kappa_2}+\pi ^2)\biggr],
\ea
and
\ba
\frac{\kappa_2}{\kappa_1}=\frac{z_0(1-c)}{2\rho},\quad \frac{s_1}{\kappa_1}=\frac{z_0(1+c)}{2},\quad 
\frac{s_1}{\kappa_2}=\frac{\rho(1+c)}{1-c}.
\ea

The soft photon emission for our kinematics  has the form
\ba
\frac{\dd\sigma_{soft}}{\dd\sigma_B}=-\frac{4\pi\alpha}{16\pi^3}
\int\frac{\dd^3k}{\omega}
\left 
(\frac{p_1}{p_1k}-\frac{p_2}{p_2k}\right 
)^2_{\omega=\sqrt{\vec{k}^2+\lambda^2}<\Delta\epsilon<<\epsilon_1\sim\epsilon_2}. \ea
Standard calculations lead to the result
\ba
\frac{\dd\sigma_{soft}}{\dd\sigma_B}=\frac{\alpha}{\pi}[(L-1)
\ln(\frac{m^2\Delta\varepsilon^2}{\lambda^2
\varepsilon_1
\varepsilon_2})
+\frac{1}{2}L^2-\frac{1}{2}\ln^2\frac{\varepsilon_1}{\varepsilon_2}-
\frac{\pi^2}{3}
+{\rm Li}_2(\frac{1-c}{2})].
\ea
The resulting contribution to the cross section from virtual and soft 
real photons does not depend on the fictitious "photon mass" $\lambda$
as well as the terms of $L^2$ type. 
It can be written in the form
\ba
\biggl(\frac{\dd\sigma}{\dd z \dd c}\biggr)_{sv}&=&\frac{\dd\sigma_{virt}+\dd\sigma_{soft}}{\dd c}\delta(z-z_0)
\\ \nonumber
&=&\frac{\alpha}{2\pi}\frac{\dd \sigma_B(p_1,\theta)}{\dd c}
\biggl[(L-1)(P_{1\Delta}+P_{2\Delta})+2K_{SV}\biggr]\delta(z-z_0),
\ea
where we have introduced the notation
\ba
P_{1\Delta}=\frac{3}{2}+2\ln\frac{\Delta\varepsilon}{\varepsilon_1}, \quad
P_{2\Delta}=\frac{3}{2}+2\ln\frac{\Delta\varepsilon}{\varepsilon_2}.
\ea
We can see that the terms proportional to the "large" logarithm $L$ have 
the form conforming with the RG prescription of the structure function.
The contribution of nonleading terms $K_{SV}$ is
\ba
\label{contrKsv}
K_{SV}=-\frac{\pi ^2}{6}+{\rm Li}_2(\frac{1-c}{2})
-\frac{1}{2}\ln ^2\frac{z_0}{\rho}+K_V.
\ea

\section{Hard collinear real photon emission contribution}
\label{Kcoll}

The dependence on the auxiliary parameter $\Delta\epsilon$ will 
be eliminated when taking into account the emission of real 
additional hard photon with 4-momentum $k$ and the
energy $\omega$ exceeding $\Delta\epsilon$.

It is convenient to consider the kinematics when this additional 
photon moves within the narrow cone of the angular size 
$m/\epsilon_1\ll\theta_0\ll1$ along the directions of the initial or
scattered electrons. The contribution of these kinematic
regions can be obtained by using the "Quasi Real Electron Method"
\cite{BFK} instead of using a general (rather cumbersome) expression
for the cross section of the double Compton (DB) scattering process \cite{MS}.

In the case when the collinear photon is emitted along the initial 
electron the result has the form:
\begin{gather}
(\frac{\dd\sigma}{\dd z \dd c})_{\vec{k}||\vec{p}_1}=
\frac{\alpha}{2\pi}\int\limits_0^{1-\frac{\Delta\epsilon}{\epsilon_1}} 
dx \frac{\dd\sigma_B}{\dd c}(xp_1,\theta)
[\frac{1+x^2}{1-x}(L_1-1)+1-x]\delta(z-t(x)),\; \\ \nonumber
L_1=\ln\frac{\theta_0^2\epsilon_1^2}{m^2}=L+\ln\frac{\theta_0^2\rho}{2z_0(1+c)}.
\end{gather}
When the photon is emitted along the scattered electron we have:
\begin{gather}
(\frac{\dd\sigma}{\dd z \dd c})_{\vec{k}||\vec{p}_2}=\frac{\alpha}{2\pi}
\frac{\dd\sigma_B}{\dd c}(p_1,\theta)\int\limits_{z(1+\frac{\Delta\epsilon}{\epsilon_2})}^{z_0}\frac{\dd t}{t}
\delta(t-z_0)[\frac{1+\frac{z^2}{t^2}}{1-\frac{z}{t}}(L_2-1)+1-\frac{z}{t}], 
\\ \nonumber 
L_2=\ln\frac{\epsilon_2^{'2}\theta_0^2}{m^2}=L+\ln\frac{\theta_0^2z^2}{2\rho(1+c)z_0},
\end{gather}
where $z=\varepsilon_2^{'}/\omega_1<z_0$ is the energy fraction of the scattered 
electron (after emission
of the collinear photon).

It is convenient to write down the contribution of the collinear kinematics
in the form
\ba
(\frac{\dd\sigma_h}{\dd z \dd c})_{coll}&=&\frac{\alpha}{2\pi}(L-1)\biggl[\int\limits_0^1\dd x\frac{1+x^2}{1-x}\theta(1-x-\Delta_1)
\frac{\dd\sigma_B(xp_1,\theta)}{\dd c}\delta(z-t(x)) \nonumber \\
&+&\int\limits_z^{z_0}\frac{\dd t}{t}\frac{1+(\frac{z}{t})^2}{1-\frac{z}{t}}\theta(1-\frac{z}{t}
-\Delta_2)\frac{\dd\sigma_B(p_1,\theta)}{\dd c}\delta(t-z_0)\biggr]+
\frac{\dd f^{(1)}}{\dd z \dd c}+\frac{\dd f^{(2)}}{\dd z \dd c},
\ea
with
\ba
\label{22}
\frac{\dd f^{(1)}}{\dd z \dd c}&=&\frac{\alpha^3}{4\rho (1-c)\omega_1^2}\biggl(\frac{2-z(1+c)}{2}
+\frac{2}{2-z(1+c)}\biggr) \\ \nonumber
&\times&\biggl[\frac{1+x^2}{1-x}\ln\frac{\rho\theta_0^2}{2z_0(1+c)}+1-x\biggr ]_{x=x_0}
\theta(1-x-\Delta_1),
\\ \nonumber 
\frac{\dd f^{(2)}}{\dd z \dd c}&=&\frac{\alpha^3}{4\rho a\omega_1^2}\biggl(\frac{1-c}{a}+\frac{a}{1-c}\biggr)
\\ \nonumber
&\times&\biggl[\frac{1+\frac{z^2}{t^2}}{1-\frac{z}{t}}
\ln\frac{z^2\theta_0^2}{2\rho(1+c)z_0}+1-\frac{z}{t}\biggr]_{t=z_0}
\theta(1-\frac{z}{t}-\Delta_2),
\quad
\Delta_{1,2}=\frac{\Delta\varepsilon}{\varepsilon_{1,2}}.
\ea
We use here the relation $\delta(z-t(x))=(2x_0^2\rho/(z^2(1-c))\delta(x-x_0)$.

Again we can see that the terms containing large logarithm $L$ 
have the  form conforming with the structure function. So our ansatz (5) is confirmed.

The auxiliary parameter 
$\theta_0$ dependence vanishes when taking into account the contribution of 
noncollinear kinematics of the additional hard photon emission (see Section \ref{appA}).

\section{Noncollinear kinematics contribution. Double Compton scattering process}
\label{conlus}

The general expression for the cross section of the DC scattering process
\begin{gather}
\gamma(k_1)+e^-(p_1) \to \gamma(k_2)+\gamma (k)+e^-(p_2),\\ \nonumber
\kappa=2kp_1,\quad \kappa'=2kp_2, \quad
\kappa_2=2k_2p_1,\quad \kappa_2^{'}=2k_2p_2,
\end{gather}
was obtained years ago by Mandl and Skyrme \cite{MS}. The expression
for the cross section presented in this paper is exact but unfortunately,
too complicated. Instead, we use the expression for the 
differential cross section calculated (by the methods of chiral
amplitudes \cite{Ber}) with the assumption that all kinematic invariants
are large compared with the electron mass squared $\kappa \sim \kappa'
\sim \kappa_i \sim  \kappa _i'\gg m^2$
\begin{gather}
\frac {\varepsilon_2\dd\sigma^{DC}_0}{\dd^3 p_2}=
\frac{1}{2!}\frac {\alpha ^3}
{2\pi ^2\kappa_1}R \dd\Phi,
\quad
\dd\Phi=\frac{\dd^3k_2}{\omega_2}\frac{\dd^3k}{\omega}
\delta^4(p_1+k_1-p_2-k_2-k),
\\ \nonumber
R=
s_1\frac {\kappa \kappa ' (\kappa ^2 +\kappa '^2) +
\kappa_1 \kappa_1 ' (\kappa_1 ^2 +\kappa_1 '^2) +
\kappa_2 \kappa_2 ' (\kappa_2 ^2 +\kappa_2 '^2)}
{\kappa \kappa ' \kappa_1 \kappa_1 ' \kappa_2 \kappa_2 ' }.
\end{gather}

The explicit expression for the contribution to the $K$ factor from hard photon 
emission $K_h$ is
\ba
\label{contrKh}
\frac{\alpha}{\pi}\frac{\dd\sigma_B}{\dd c} K_h=\frac{\dd\sigma^{DC}_{\theta_0}}{\dd z \dd c}+
\frac{\dd f^{(1)}}{\dd z \dd c}+\frac{\dd f^{(2)}}{\dd z \dd c}.
\ea
with 
\ba
\frac{\dd\sigma^{DC}_{\theta_0}}{\dd z \dd c}=\frac{\alpha^3z}{2!4\pi\rho}
\int R \dd\Phi,
\ea
and the phase volume $\dd\Phi$ is restricted by the conditions $\omega,\omega_2>\Delta\epsilon$ 
and the requirement that the angles between 3-vectors $\vec{k}_2,\vec{k}$ and 3-vectors
$\vec{p}_1$, $\vec{p}_2$ exceed $\theta_0$.

The values of $K_h$ calculated numerically are given in Tables 1,7.  We show 
numerically and  analytically (see Appendix A)the independence of $K_h$ 
on the auxiliary parameters $\theta_0,\Delta\varepsilon$.

The cross section of the DC scattering process in an inclusive experimental set-up
with the leading 
logarithmic approximation in terms of structure functions has the form
\ba
\dd\sigma^{DC}(p_1,k_1;p_2,k,k_2)=
\int\limits_0^1 \dd x D(x,L)\int\limits_z^1 D(\frac{z}{t})\frac{\dd t}{t}
\dd\sigma_0^{DC}(xp_1,k_1;tp_2/z,k,k_2),
\ea
with the structure functions given above and
\ba
\dd\sigma^{DC}_0(p_1,k_1;p_2,k,k_2)= 
\frac {\alpha ^3}{4\pi ^2\kappa_1}R
\frac{\dd^3k_2\dd^3k\dd^3p_2}{\omega_2\omega 
\epsilon_2}\delta^4(p_1+k_1-p_2-k_2-k).
\ea

\section{Conclusion}
\label{conc}
\begin{figure} 
\includegraphics[scale=0.6]{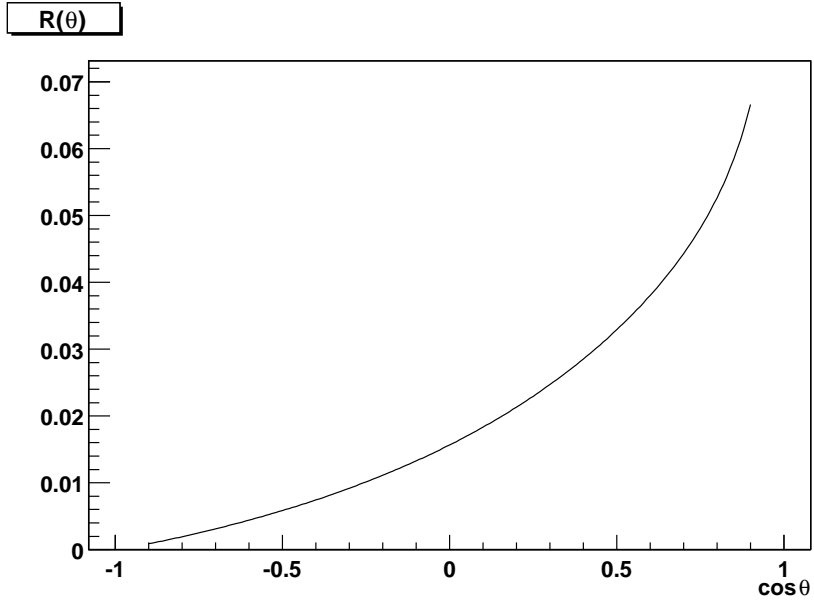}
\caption{ The leading order radiative corrections as $\cos\theta$
distribution (see formulae (\ref{intz})).}
\label{fig1}
\end{figure}

The characteristic form "reverse radiative tail" (see Tables 2,4) of the 
differential cross section
on the energy fraction $z$ can be reproduced if one uses the "smoothed"
expression for nonsinglet structure functions which includes the virtual
electron pair production \cite{KF}
\ba
\label{smoothed}
D(x,L)= \frac{\beta}{2}(1-x)^{\beta/2-1}(1+\frac{3}{8}\beta)-\frac{\beta}{4}(1+x)+
O(\beta^2),\quad \beta=\frac{2\alpha}{\pi}(L-1), \\ \nonumber
O(\beta^2)=\frac{\beta}{2}(1-x)^{\beta/2-1}(-\frac{1}{48}\beta^2(\frac{1}{3}L
+\pi^2-\frac{47}{8}))+\frac{1}{32}\beta^2(-4(1+x)\log(1-x)\\ \nonumber
-\frac{1+3x^2}{1-x}\log x-5-x).
\ea 
In Figure (\ref{fig1}) we put the magnitude of RC in the leading
approximation
\ba
\label{intz}
R(\theta)=\biggl(\frac{\dd\sigma_B}{\dd c}\biggr)^{-1}
\biggl(\int\dd z\frac{\dd\sigma}{\dd z \dd c}-\frac{\dd\sigma_B}{\dd c}\biggr).
\ea

The results cited above imply the experimental set-up without additional
$e^+e^-,\mu^+\mu^-,\pi^+\pi^-$ real pairs in the final state.

The accuracy of the formulae given above is determined by the order of magnitude of the terms 
omitted (see (2)) compared to the terms of order of unity, i.e., is of the order of $0.1\%$
for typical experimental conditions. In particular, it is the reason
why we omit the evolution effect of the $K$-factor terms.

The numerical value of $K_h$, leading contributions,  and the Born cross-section 
for different kinematic regions
are presented as a functions of $z,c$ in  Tables 1-5,7.

\begin{center}
\begin{tabular} {||c||c|c|c|c|c|c|c|c|c||}  \hline\hline
$z\backslash \cos\theta$ & -0.8&-0.6&-0.4&-0.2& 0.0&0.2&0.4&0.6 &0.8 \\ \hline\hline
0.1 &-2.82 &-2.61&-2.39&-2.19&-2.09&-1.89&-1.87&-2.06&-2.75 \\ \hline
0.2 & -2.77&-2.47&-2.17&-1.90&-1.65&-1.46&-1.39&-1.56&-2.30 \\ \hline
0.3 & -3.43&-2.98&-2.55&-2.14&-1.77&-1.47 &-1.30 &-1.38&-2.13 \\ \hline
0.4 & -4.96&-3.87&-3.23&-2.65&-2.13 &-1.67 &-1.34 &-1.30&-2.02 \\ \hline\hline
\end{tabular}
\end{center}
Table 1: The value of $K_h$ as a function of $z$, $\cos\theta$
(calculated for $\rho=0.4$).

\begin{center}
\begin{tabular} {||c||c|c|c|c|c|c|c|c|c||}  \hline\hline
$z\backslash \cos\theta$ & -0.8&-0.6&-0.4&-0.2& 0.0&0.2&0.4&0.6 &0.8 \\ \hline\hline
0.1 & 0.211 & 0.237 &0.265&0.299&0.345&0.413 &0.526 &0.754&1.450 \\ \hline
0.2 & 0.337 & 0.357 &0.378&0.405&0.445&0.508 &0.618 &0.850&1.576 \\ \hline
0.3 & 0.703 & 0.669 &0.643&0.634&0.644&0.685 &0.782 &1.013&1.784 \\ \hline
0.4 & 3.883 & 2.153 &1.554&1.264&1.113 &1.054&1.090 &1.296&2.122 \\ \hline\hline
\end{tabular}
\end{center}
Table 2: The value of $\omega_1^2/\alpha^2\dd\sigma/(\dd c\dd z)$ 
(leading contribution, first term in the right-hand side of the 
master formula (\ref{master}) ) as a function of $z$, $\cos\theta$
(calculated for $\rho=0.4$, $\omega_1=5$ GeV).

\begin{center}
\begin{tabular} {||c||c|c|c|c|c|c|c|c|c||}  \hline\hline
$\cos\theta$ & -0.8&-0.6&-0.4&-0.2& 0.0&0.2&0.4&0.6 &0.8 \\ \hline\hline
$\frac{\omega_1^2}{\alpha^2}\frac{\dd\sigma_B}{\dd c}$ & 1.779 & 2.038 
&2.365&2.796&3.389&4.266&5.721 &8.669&17.881 \\ \hline\hline
\end{tabular}
\end{center}
Table 3: Born cross section (\ref{born}) (without factor  $\alpha^2/\omega_1^2$) for $\rho=0.4$.

\section{Appendix A}
\label{appA}
Performing the integration over $k_2$ of the phase volume
\ba
d\Phi=\frac{d^3k}{\omega}\frac{d^3k_2}{\omega_2}\delta^4(Q-k-k_2),\quad  Q=p_1+k_1-p_2,
\ea
we can put it in the form
\ba
d\Phi=\frac{\omega d \omega}{\omega_1^2}\frac{2d c_1 d c_2}{\sqrt{D}}
\delta[2\rho-\rho z (1+c)-z(1-c)-\frac{\omega}{\omega_1}
(\rho(1-c_1)-z(1-c_2)+1+c_1)],
\ea
with  $D=1-c_1^2-c_2^2-c^2-2cc_1c_2$; $c_1,c_2$ are the cosines of the angles between $\vec{k}$ and $\vec{p}_1,\vec{p}_2$
respectively.

For collinear kinematics the following relation can be useful:

1. $k\approx (1-x)p_1$
\begin{gather}
R_1=R|_{\vec{k}||\vec{p}_1}
=\biggl(\frac{2x\rho}{z(1-c)}+\frac{z(1-c)}{2x\rho}\biggr)\frac{1+x^2}{(1-x)^2}\frac{1}{2\rho^2(1-c_1)x\omega_1^2},
\nonumber \\ \nonumber
\dd \Phi_1=\dd \Phi|_{\vec{k}||\vec{p}_1}=2\frac{\dd^3 k}{\omega}\delta((xp_1+k_1-p_2)^2)
=2\pi\frac{\rho(1-x)\dd x\dd c_1}{2-z(1+c)}\delta(x-x_0),
\\ 
\frac{\dd\sigma^1_{h}}{\dd z \dd c}=\frac{\alpha^3z}{2! 4\pi\rho}\int R_1\dd \Phi_1
=\frac{\alpha^3}{4\rho\omega_1^2(1-c)}\frac{1+x_0^2}{1-x_0}\biggl(\frac{2x_0\rho}{z(1-c)}+\frac{z(1-c)}{2x_0\rho}\biggr)
\ln(\frac{4}{\theta_0^2}).
\label{kdf11}
\end{gather}
In the last equation we take into account the same contribution from
the region  $k_2\approx(1-x)p_1$  

2. For the case $k\approx(t/z-1)p_2$ we obtain
\begin{gather}
R_2=R|_{\vec{k}||\vec{p}_2}
=\biggl(\frac{2x\rho}{z(1-c)}+\frac{z(1-c)}{2x\rho}\biggr)\frac{1+x^2}{(1-x)^2}\frac{1}{2\rho^2(1-c_1)x\omega_1^2},
\\ \nonumber
\dd \Phi_2=\dd \Phi|_{\vec{k}||\vec{p}_2}=2\frac{\dd^3 k}{\omega}\delta((xp_1+k_1-p_2)^2)
=2\pi\frac{\rho(1-x)\dd x\dd c_1}{2-z(1+c)}\delta(x-x_0),
\end{gather}
So the contribution of the case $\vec{k}||\vec{p}_2$ ($\vec{k}_2||\vec{p}_2$) has the form
\begin{gather}
\label{k2}
\frac{\dd\sigma^2_{h}}{\dd z \dd c}=\frac{\alpha^3z}{2! 4\rho\omega_1^2}\int R_2\dd \Phi_2
=\frac{\alpha^3}{4\rho a\omega_1^2}(\frac{1-c}{a}+\frac{a}{1-c})
\frac{1+\frac{z^2}{t^2}}{1-\frac{z}{t}}\ln(\frac{4}{\theta_0^2}).
\end{gather}
Comparing formulae (\ref{kdf11},\ref{k2}) with (\ref{22}) we can see explicit
cancellation of the $\theta_0$ dependence.

\section{Appendix B}
\label{secK}
Here we put the different case of kinematic region for $\rho$, $z$

All the above formulae were considered for the case $\rho< 1$, and
the possible region for the variable $z$ was determined by the 
equation $x_0<1$
\begin{gather}
z\le\frac{2\rho}{1-c+\rho(1+c)},
\end{gather} 
which means that the low boundary of integration in formula (\ref{master})
is less than $1$.
In the case of $\rho>1$  it is convenient to put the new variable
\begin{gather}
\eta=\frac{\omega_1}{\varepsilon_1}, \quad y=\frac{\varepsilon_2^{'}}{\varepsilon_1},\quad
y_0=\frac{\varepsilon_2}{\varepsilon_1}=\frac{2\eta}{1+c+\eta(1-c)}, \quad
\eta<1.
\end{gather}
The master equation (\ref{master}) for the case $\rho>1$ (or $\eta<1$) reads
\begin{gather}
\label{masterII}
\frac{\dd\tilde{\sigma}}{\dd y \dd c}(p_1,p_2)
=\int\limits_{\tilde{x}_0}^1\frac{\dd x}{\tilde{t}(x)}D(x,\tilde{L})
\frac{\dd\tilde{\sigma}_B(xp_1,\theta)}{\dd c}D(\frac{y}{\tilde{t}(x)},\tilde{L})+
\frac{\alpha}{\pi}\frac{\dd\tilde{\sigma}_B(p_1,\theta)}
{\dd c}\biggl[\tilde{K}_{SV}\delta(y-y_0)+\tilde{K}_h\biggr],\\ \nonumber
\tilde{x}_0=\frac{y\eta(1-c)}{2\eta-y(1+c)}, 
\quad \tilde{L}=\ln\frac{2\varepsilon_1^2 y_0(1+c)}{m^2},\quad
\tilde{t}(x)=\frac{2\eta x}{x(1+c)+\eta(1-c)}.
\end{gather}
with the possible values for energy fraction of the scattered electron $y$($\tilde{x}_0<1$) :
$y\le y_0$.
The Born cross section (\ref{born},\ref{bornx}) and formulae for hard photon emission,
$\tilde{K}_{SV}$, $\tilde{K}_h$ for the case $\rho>1$ appear just by appropriate exchange $\rho\to \eta^{-1}$:
\begin{gather}
\label{borneta}
\frac{\dd\tilde{\sigma}_B(xp_1,\theta)}{\dd c}=\frac{\pi\alpha^2}{\varepsilon_1^2}
\frac{1}{(\eta(1-c)+ x(1+c))^2}
\biggl(\frac{\eta(1-c)}{\eta(1-c)+x(1+c)}+\frac{\eta(1-c)+ x(1+c)}{\eta(1-c)}\biggr).
\end{gather}

\begin{center}
\begin{tabular} {||c||c|c|c|c|c|c|c|c||}  \hline\hline
$y\backslash \cos\theta$ & -0.8&-0.6&-0.4&-0.2& 0.0&0.2&0.4&0.6  \\ \hline\hline
0.05 & 9.658 & 11.110  &13.626&17.513&23.678&34.116  &53.669 &98.208     \\ \hline
0.10 & 11.350 & 15.024 &22.633&39.297&86.017&        &       &   \\ \hline
0.15 & 13.839 & 23.190 &56.097&      &      &        &       &   \\ \hline
0.20 & 17.735 & 45.672 &      &      &      &        &       &   \\ \hline
0.25 & 24.303 &        &      &      &      &        &       &   \\ \hline\hline 
\end{tabular}
\end{center}
Table 4: The value of $\varepsilon_1^2/\alpha^2\dd\tilde{\sigma}/(\dd c\dd y)$ 
(leading contribution, first term in the right-hand side of the master 
formula (\ref{masterII}) ) as a function of $z$, $\cos\theta$
(calculated for $\omega_1=400$ Mev, $\varepsilon_1=6$ GeV).

\begin{center}
\begin{tabular} {||c||c|c|c|c|c|c|c|c|c||}  \hline\hline
$\cos\theta$ & -0.8&-0.6&-0.4&-0.2& 0.0&0.2&0.4&0.6 &0.8 \\ \hline\hline
$\frac{\varepsilon_1^2}{\alpha^2}\frac{\dd\tilde{\sigma}_B}{\dd c}$ 
& 93.317 & 60.706 &49.428&44.994&44.351&47.084&54.584&72.444&129.944 \\ \hline\hline
\end{tabular}
\end{center}
Table 5: Born cross section (\ref{borneta}) (without factor  $\alpha^2/\omega_1^2$) 
for $\omega_1=400$ Mev, $\varepsilon_1=6$ GeV.

Large amounts of the leading contribution near the kinematic bound can be understood
as  manifestation of the $\delta(y-y_0)$ character of the differential cross section.
The $y_0$, $z_0$ dependence is given in Table 6.

\begin{center}
\begin{tabular} {||c||c|c|c|c|c|c|c|c|c||}  \hline\hline
$\cos\theta$ & -0.8&-0.6&-0.4&-0.2& 0.0&0.2&0.4&0.6 &0.8 \\ \hline\hline
$y_0$ & 0.417 & 0.263 &0.192&0.152&0.125&0.106&0.093&0.082&0.074 \\ \hline
$z_0$ & 0.423 & 0.455 &0.489&0.526&0.571&0.625&0.690&0.769&0.870 \\ \hline\hline
\end{tabular}
\end{center}
Table 6: The value of $y_0$, $z_0$ as a function of $c$ for $\eta=0.064$
and $\rho=0.4$.

\begin{center}
\begin{tabular} {||c||c|c|c|c|c|c|c|c|c||}  \hline\hline
$y\backslash \cos\theta$ & -0.8&-0.6&-0.4&-0.2& 0.0&0.2&0.4&0.6 &0.8 \\ \hline\hline
0.05 & 0.70 & -1.97  &-7.41&-15.54&-26.90&-42.70 &-65.40 &-100.64&-166.21 \\ \hline
0.10 & 0.36 & -3.20 & -9.85 &-18.38 &-18.35 &    &       & &  \\ \hline
0.15 & 0.03 & -3.38 & -1.34 &        &      &        &       & &  \\ \hline
0.20 & -0.20 & 0.29 &      &      &      &        &       & &  \\ \hline
0.25 & -0.25 &      &      &      &      &        &       & &  \\ \hline\hline 
\end{tabular}
\end{center}
Table 7: The value of $\tilde{K}_h$  as a function of $y$, $\cos\theta$
(calculated for $\eta=0.064$).

\section{Acknowledgements}

Two of us (EAK) and (VVB) are grateful to RFFI, grant 03-02-17077, for supporting this work.
We are grateful to Sergej Gevorkyan for taking part at the beginning of this work
and S. Dittmaier for reminding us of the valuable set of previously published papers.

\end{document}